\documentclass[aps,prb,twocolumn,showpacs]{revtex4}
\usepackage{bm}

\usepackage{amsmath}
\usepackage{amssymb}
\usepackage{graphicx}
\usepackage{psfrag}

\newcommand{\beq}{\begin{equation}}
\newcommand{\eeq}{\end{equation}}
\newcommand{\beqar}{\begin{eqnarray*}}
\newcommand{\eeqar}{\end{eqnarray*}}

\newcommand{\ua}{\uparrow}
\newcommand{\da}{\downarrow}

\newcommand{\dt}{{\text d}}
\newcommand{\etxt}{{\text e}}

\newcommand{\itxt}{{\text i}}

\newcommand{\mtxt}{{\text m}}

\newcommand{\ttxt}{{\text t}}

\newcommand{\ph}{\hat{p}}

\newcommand{\Hh}{\hat{H}}

\newcommand{\sigb}{{\mbox{\boldmath{$\sigma$}}}}

\newcommand{\pd}{\partial}

\newcommand{\s}{{\bf s}}

\newcommand{\Bb}{{\bf B}}

\newcommand{\dg}{\dagger}

\newcommand{\lan}{\langle}
\newcommand{\ran}{\rangle}
\newcommand{\om}{\omega}

\newcommand{\al}{\alpha}

\newcommand{\de}{\delta}
\newcommand{\De}{\Delta}

\newcommand{\sig}{\sigma}

\newcommand{\ka}{\varkappa}
\newcommand{\vphi}{\varphi}

\newcommand{\e}{\epsilon}

\newcommand{\lt}{\left}
\newcommand{\rt}{\right}
\begin{document}
\title{Interaction-enhanced magnetically ordered insulating state at the edge \\
of a two-dimensional topological insulator}
\author{Maxim Kharitonov}

\address{
Materials Science Division, Argonne National Laboratory, Argonne, Illinois 60439, USA\\
Center for Materials Theory, Rutgers University, Piscataway, New Jersey 08854, USA
}
\date{\today}
\begin{abstract}

We develop a theory of the correlated  magnetically ordered insulating state
at the edge of a two-dimensional topological insulator.
We demonstrate that the gapped spin-polarized state, induced by the application of the magnetic field $B$,
is naturally facilitated by electron interactions, which drive the critical easy-plane ferromagnetic correlations in the helical liquid.
As the key manifestation, the gap $\De$ in the spectrum of collective excitations, which carry both spin and charge,
is enhanced and exhibits a scaling dependence $\De \propto B^{1/(2-K)}$,
controlled by the Luttinger liquid parameter $K$. This scaling dependence could be probed through
the activation behavior $G \sim (e^2/h) \exp(- \De/T)$ of the longitudinal conductance of a Hall-bar device at lower temperatures,
providing a straightforward way to extract the parameter $K$ experimentally.
Our findings thus suggest that the signatures of the interaction-driven quantum criticality of the helical liquid could be revealed already
in a standard Hall-bar measurement.

\end{abstract}
\maketitle

\section{Introduction}
\begin{figure}
$\mbox{\includegraphics[width=.27\textwidth]{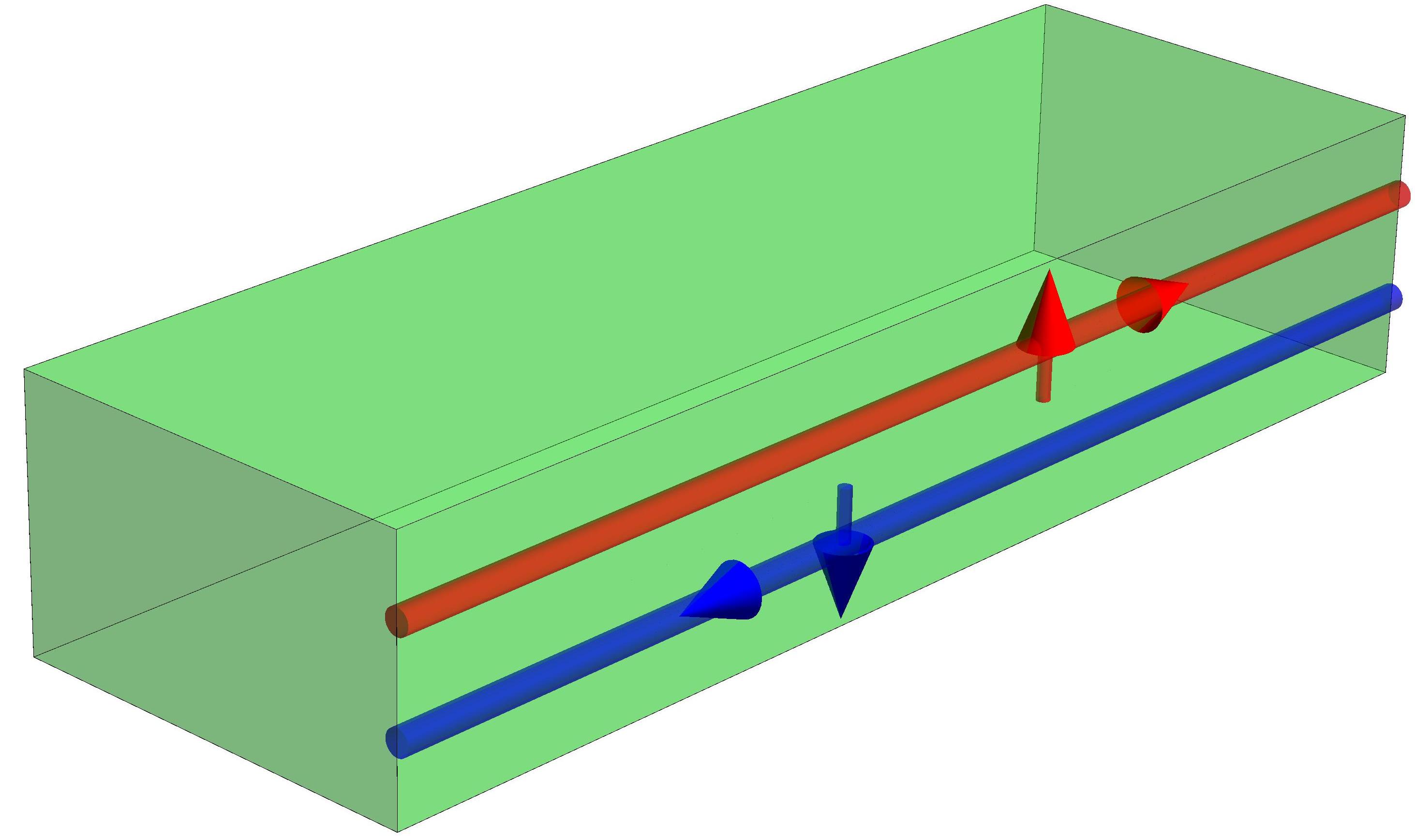}}\mbox{ }
\mbox{\includegraphics[width=.20\textwidth]{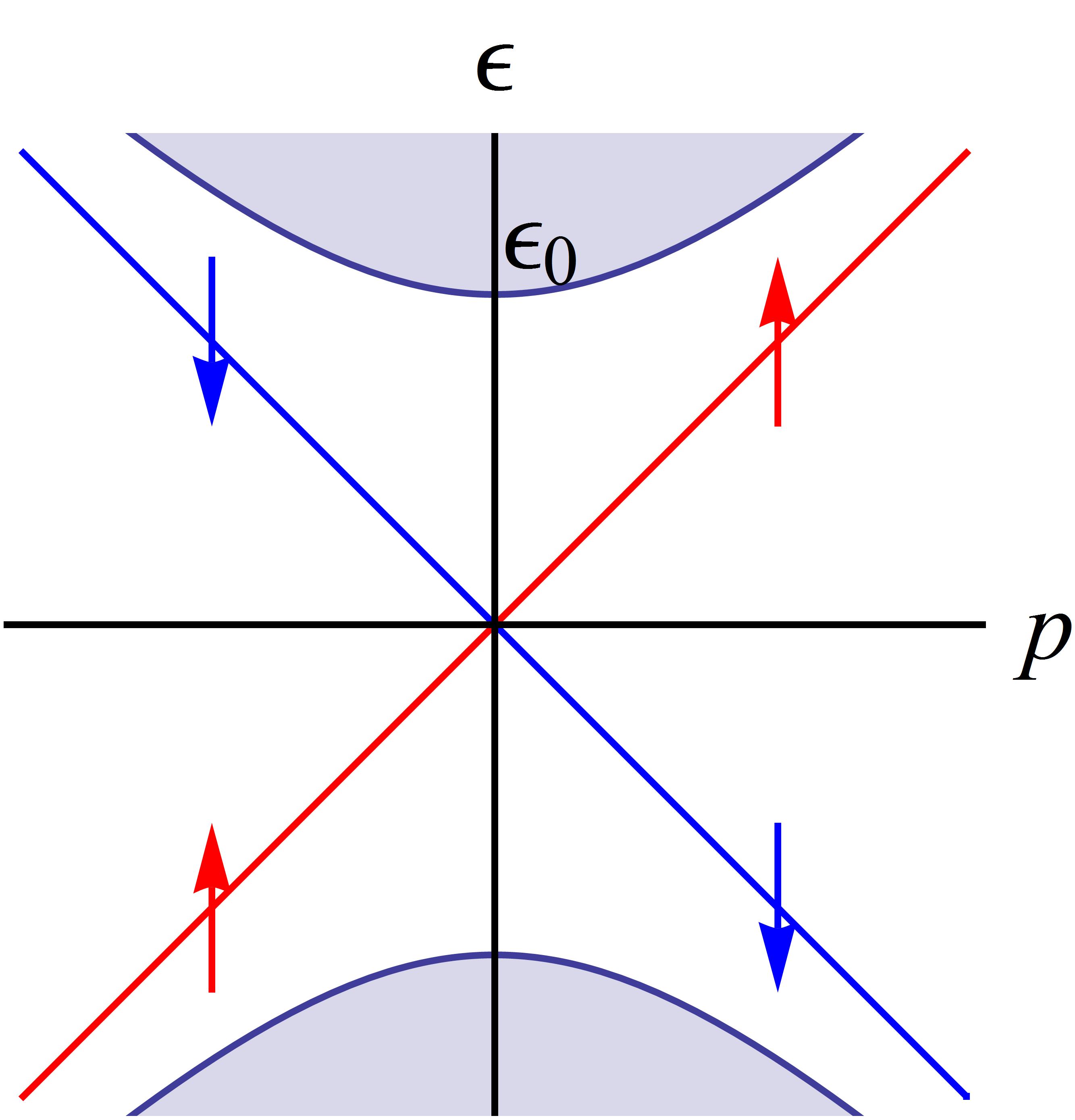}}$
\caption{(Color online)
Helical edge states of a 2D topological insulator. (Left) The states propagating in the opposite directions
have opposite spin projections on the direction perpendicular to the plane of the sample.
(Right) In the absence of the magnetic field the counter-propagating states are gapless.
Shaded regions depict the continuum of the extended bulk states with the insulating gap $\e_0$.
}
\label{fig:2DTI}
\end{figure}

Topological insulators~\cite{TItheor0,TItheor01,TItheor1,TItheor2,TItheor3,TItheor4,TItheor5,Roy,TIexp1,TIexp2,TIexp3,TIexp4,HgTetheor,HgTeexp,HgTerev}
form a new class of materials with nontrivial band structure caused by spin-orbit interactions.
The key physical feature that distinguishes a topological insulator (TI) from a conventional, nontopological, one is the presence of gapless
surface or edge electron states.
The edge of a two-dimensional (2D) topological insulator~\cite{TItheor0,TItheor01,TItheor3,TItheor4,HgTetheor}
supports two branches of gapless counter-propagating helical states
with opposite spin projections on the axis perpendicular to the plane of the sample (Fig.~\ref{fig:2DTI}).
Protected by the time-reversal symmetry against single-particle nonmagnetic backscattering~\cite{HL1,HL2},
these edge modes serve as nearly ideal conducting channels that give rise to the quantum spin Hall effect.
So far, a 2D topological insulator was realized in HgTe-CdTe quantum wells,
which was first predicted theoretically~\cite{HgTetheor}  and shortly after confirmed experimentally~\cite{HgTeexp,HgTerev}.

Interactions between electrons in the counter-propagating states
lead to a one-dimensional helical Luttinger liquid (LL) phase~\cite{HL1,HL2,sigrho,TIKondo,HLpoint1,HLpoint2,HLpoint3,2HL,Das,Tanaka,Schmidt,Dolcetto},
which hosts a number of remarkable physical properties,
such as quantum criticality, bonding of the spin and charge degrees of freedom, and charge fractionalization.
However, interaction effects in a LL are generally known to be quite elusive to experimental probes.
In particular, for negligible single-particle backscattering,
the longitudinal conductance $e^2/h$ of a LL remains essentially unaffected by the interactions~\cite{MS,SafiSchulz}.
In a helical LL, this holds as long as time-reversal symmetry is preserved and the system remains gapless.
Probing interactions in this regime by a transport measurement generally requires creating a tunneling setup
of some kind~\cite{HLpoint1,HLpoint2,HLpoint3,2HL,Das,Dolcetto}.

In this paper, we demonstrate that electron interactions
in a helical liquid
reveal themselves in an interesting fashion once the time-reversal symmetry is broken by the application of an external magnetic field.
Indeed, on the one hand, in the noninteracting picture,
the magnetic field couples the counter-propagating edge states, opens a gap in the single-particle spectrum,
and spin-polarizes the edge.
On the other hand,
in the absence of the magnetic field,
interactions in a helical LL result in a tendency towards easy-plane ferromagnetism,
manifested in a critical power-law decay of the spin correlations.
Therefore, once the magnetic field is applied,
one can naturally expect electron interactions to
facilitate the formation of the spin-polarized state.

The present paper is devoted to the theory of this correlated magnetically ordered insulating state,
induced by the magnetic field and enhanced by the interactions, at the edge of a 2D topological insulator.
Our key finding is that the gap $\De$ in the spectrum of collective excitations is enhanced by the interactions
and exhibits a critical scaling dependence $\De \propto B^{1/(2-K)}$ on the magnetic field $B$.
Its exponent is controlled by the LL parameter $K$, which characterizes the interaction strength.
Crucially, this critical scaling should reveal itself
in the low-temperature activation behavior $G \sim (e^2/h) \exp(- \De/T)$
of the longitudinal conductance of a Hall-bar device,
which allows one to extract the LL parameter $K$ and infer about the strength of interactions in a real system.
Our work suggests that the interaction-driven quantum criticality of the helical liquid at the edge of a 2D topological
insulator could be accessed already via a standard Hall-bar measurement.

The suppression of the longitudinal conductance with the applied magnetic field was already observed experimentally in HgTe
quantum wells~\cite{HgTeexp,HgTerev}.
However, two factors preclude direct comparison of the present prediction with that data:
(i) the magnetic-field data were provided for a large sample of size $20\times13 \mu\mtxt^2$,
for which backscattering was substantial;
(ii) the temperature dependence of the conductance, necessary to extract the transport gap $\De$, was not provided.

\section{Model and Hamiltonian}
The effective low-energy Hamiltonian for the interacting electrons in the counter-propagating edge states of a 2D topological insulator
in the presence of a magnetic field~\cite{HgTerev,sigrho}
may be written down in the helical basis of right-moving (with respect to the $x$ direction along the edge) spin-up ($\ua$) and left-moving
spin-down ($\da$) states as
\beq
    \Hh=\Hh_0 + \Hh_\mtxt+\Hh_\itxt
    ,\mbox{ } \Hh_0 = \int \dt x \, \psi^\dg(x) \,v \ph \, \sig_z \, \psi(x),
\label{eq:H}
\eeq
\beq
    \Hh_\mtxt = -\De_0\int \dt x \, \psi^\dg(x) (\sig_x \cos \vphi_0 +\sig_y \sin  \vphi_0) \psi(x),
\label{eq:Hm}
\eeq
\beq
\Hh_\itxt=\frac{1}{2} \int \dt x \, \dt x' \,  \psi^\dg_\sig (x) \psi^\dg_{\sig'} (x') V(x-x') \psi_{\sig'} (x') \psi_{\sig} (x).
\label{eq:Hi}
\eeq
Here, $\psi= (\psi_\ua, \psi_\da)^\ttxt$ is the two-component fermionic field operator, $\ph=-\itxt  \hbar \pd_x$, and $\sig_x, \sig_y, \sig_z$ are the Pauli matrices in the helical basis. The part
$\Hh_\mtxt$ describes the effect of the external magnetic field.
For the in-plane orientation, $\Bb=B (\cos \vphi_0,\sin \vphi_0,0)$, only the Zeeman effect is present, whereas the orbital effect vanishes;
the angle $\vphi_0$ correspond to the direction of the field in the $xy$ plane of the 2D sample and the gap 
is given by the Zeeman energy $\De_{0 \parallel} \sim \mu_B B$.
In case of the perpendicular orientation of the field, $\Bb=(0,0,B)$, the Zeeman effect does not affect the dynamics
and only the orbital effect remains.
The orbital effect of the perpendicular field is estimated~\cite{HgTerev}
to be stronger than the in-plane Zeeman effect, $\De_{0 \perp} \sim 10  \De_{0 \parallel}$;
$\De_{0\parallel} \approx 3 \text{K}$ and $\De_{0\perp} \approx 30 \text{K}$ at $B=1 \text{T}$.
For arbitrary field orientation, the single-particle gap $\De_0$ scales linearly with the magnetic field, $\De_0 \propto B$.

We consider the case of Coulomb interactions,  $V(x)=e^2_*/|x|$ in Eq.~(\ref{eq:Hi}),
possibly screened  by the nearby metallic electrodes beyond some length $l_s$;
the charge $e_*=e/\sqrt{\ka}$ incorporates the effects of screening by the dielectric environment.
This allows us to consider both unscreened and screened interactions, the latter modeling  practically any finite-range interactions.
The short-scale spatial cutoff $\al$ of the theory [Eqs.(\ref{eq:H}), (\ref{eq:Hm}), and (\ref{eq:Hi})] and of the potential $V(x)$  is set by the decay scale of the edge states into the bulk.
For simplicity, it is assumed that the chemical potential is exactly at  the branch crossing $\e=0$ of the unperturbed edge spectrum
$\e_p = \pm v p$, where the correlation effects are strongest. This can be achieved by tuning the gate voltage to the minimum of the longitudinal conductance.

The Hamiltonian $\Hh$ [Eqs.~(\ref{eq:H}), (\ref{eq:Hm}), and  (\ref{eq:Hi})] describes one-dimensional interacting Dirac fermions, which are massive in the presence of the magnetic field; for point interactions, this is known as the Thirring model~\cite{GNT,Giamarchi}.
This fermionic model can be mapped a bosonic one by mean of the bosonization procedure~\cite{GNT,Giamarchi}.
One relates the fermion fields $\psi_{\ua,\da}(x)$ of the right and left movers to the bosonic ones $\vphi_{\ua,\da}(x)$ as
\beq
    \psi_{\ua,\da}(x) = \frac{1}{\sqrt{2 \pi \al}} \etxt^{\pm \itxt \vphi_{\ua,\da}(x)},
\label{eq:psiphi}
\eeq
where the Klein factors are omitted.
The operators
$
    \vphi(x)=\frac{1}{2}[\vphi_\ua(x)+\vphi_\da(x)]
\mbox{ and }
    \theta(x)=\frac{1}{2}[\vphi_\ua(x)-\vphi_\da(x)]
$
satisfy the canonical (up to a coefficient) commutation relations
$
    [\vphi(x), \pd_{x'}\theta(x')] =- \itxt \pi \de(x-x')
$
and are related to the coordinate and momentum variables of the collective excitations.
In terms of $\vphi(x)$ and $\theta(x)$, the Hamiltonian $\Hh$ [Eqs.~(\ref{eq:H}), (\ref{eq:Hm}), and (\ref{eq:Hi})]
can be expressed as
\beq
   \Hh_{0} = \frac{\hbar }{2\pi} \int \dt x \, v[ (\pd_x \theta)^2 + (\pd_x \vphi )^2],
\label{eq:H0b}
\eeq
\beq
    \Hh_\mtxt = - \frac{\De_0}{\pi \al} \int \dt x \, \cos[2\vphi(x)+\vphi_0],
\label{eq:Hmb}
\eeq
\beq
    \Hh_{\itxt} = \frac{1}{2\pi^2} \int \dt x \, \dt x' \, \pd_x \vphi (x) V(x-x') \pd_{x'} \vphi(x').
\label{eq:Hib}
\eeq
The Hamiltonian~(\ref{eq:H0b}), (\ref{eq:Hib}), and (\ref{eq:Hmb})
describes the dynamics of the collective edge excitations
of a 2D topological insulator in the presence of a magnetic field.
This is the sine-Gordon model~\cite{GNT,Giamarchi} for point interactions
and its nonlocal generalization for finite-range interactions.
Below we analyze the properties of this model.

\begin{figure}
\includegraphics[width=.48\textwidth]{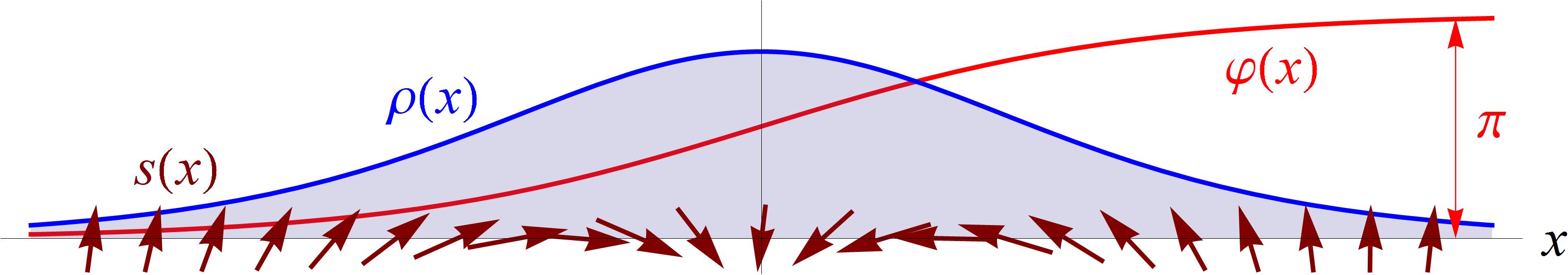}
\caption{
(Color online)
Collective spin-charge excitations of the edge of a 2D topological insulator. Excitations
are described by the  phase variable $\vphi(x)$, which determines both the in-plane spin polarization [Eq.~(\ref{eq:sxsy})] and charge density [Eq.~(\ref{eq:rho})].
As a specific illustrative example,
a kink of height $\pi$ in $\vphi(x)$ rotates the spin polarization in the $xy$ plane of the sample by $2 \pi$ and accumulates a unit charge
in the region of variation of $\vphi(x)$.}
\label{fig:sigrho}
\end{figure}

\section{Collective spin-charge excitations}
To visualize the collective excitations described by Eqs.~(\ref{eq:H0b}), (\ref{eq:Hmb}), and (\ref{eq:Hib}),
let us  link the fields $\vphi(x)$  and $\theta(x)$ to the physical observables.
From the relation (\ref{eq:psiphi}), one obtains
\beq
    \lt( \begin{array}{c}s_{x}(x) \\ s_{y}(x)\end{array}\rt) =
        \frac{1}{2 \pi \al} \lt( \begin{array}{c} \cos(-2\vphi(x)) \\ \sin(-2 \vphi(x)) \end{array}\rt)
\label{eq:sxsy}
\eeq
for  the $x$ and $y$ components of the spin density operator $\s(x) = \psi^\dg_\sig (x) \sigb_{\sig\sig'} \psi_{\sig'} (x)$  (defined without 1/2 factor)
and
\beq
    s_{z}(x) =
    \frac{1}{\pi} \pd_x \theta(x), \mbox{ }
    \rho(x)= \frac{1}{\pi} \pd_x \vphi(x)
\label{eq:rho}
\eeq
for the $z$ component of the spin density and the particle density $\rho(x)=\psi^\dg_\sig (x)  \psi_{\sig} (x)$
operators.
As seen from Eq.~(\ref{eq:sxsy}), the angle $-2 \vphi(x)$ corresponds to the direction of the spin polarization in the $xy$ plane and
the field $\vphi(x)$ is thus directly related to the spin degrees of freedom.
At the same time, according to Eq.~(\ref{eq:rho}), the charge density is determined by the gradient of $\vphi(x)$.
Therefore, the collective excitations carry {\em both} charge and spin, which is a direct consequence of
the coupling between the spin and orbital degrees of freedom in the single-particle states.
As a specific illustrative example of this property,
a kink of height $\pi$ in $\vphi(x)$ rotates the spin polarization in the $xy$ plane by $2\pi$ and  simultaneously
accumulates a unit charge in the region of variation of $\vphi(x)$, Fig.~\ref{fig:sigrho}.
It was suggested in Ref.~\cite{sigrho} to exploit this bonding of spin and charge degrees of freedom
to observe charge fractionalization effects in domain-wall structures with inhomogeneous magnetization.

\section{Gapless helical liquid at $B=0$}
Let us first consider the system in the absence of the magnetic field, $\Hh_\mtxt=0$, when the edge is in the helical LL phase,
and obtain the excitation spectrum and basic correlations.
The calculations can be conveniently performed in the Langrange finite-temperature formalism.
From Eqs.~(\ref{eq:H0b}) and (\ref{eq:Hib}), the action for the Fourier transformation
$  \vphi(\om_n,q) = \int_0^{\hbar/T} \dt \tau \int \dt x \,
    \etxt^{\itxt \om_n\tau - \itxt q x} \vphi(\tau,x)$
    ($\hbar \om_n=2\pi T n$,  $n \in \mathbb{Z}$) of the
phase field
takes the form
\beq
    S_0[\vphi]+S_\itxt[\vphi]= T \sum_{\om_n} \int \frac{\dt q}{2\pi}  \lt(\frac{1}{u_q} \om_n^2 + u_q q^2 \rt) \frac{|\vphi(\om_n,q)|^2}{2 \pi K_q}.
\label{eq:S}
\eeq
The momentum-dependent velocity  $u_q$ and
LL interaction parameter $K_q$ are given by
\beq
        u_q/v =1/K_q
        =
        \sqrt{1+ V(q)/(\pi \hbar v)}
        =  \sqrt{r_s \ln [1/(q_* \al_*)] },
\label{eq:uK}
\eeq
where $V(q)=2 e_*^2 \ln [1/(q_* \al)]$ is the Fourier transform of the 
potential $V(x)$,
$r_s=2 e^2_* /(\pi \hbar v)$ is the Coulomb parameter, $q_*= \max( |q| , 1/l_s)$,  and $\al_* \sim \al \etxt^{-1/r_s}$.

From Eqs.~(\ref{eq:S}) and (\ref{eq:uK}), one obtains
the excitation spectrum
$    \om(q)= u_q |q|
$
of the collective edge excitations of a 2D topological insulator.
For unscreened Coulomb interactions $V(q)= 2 e_*^2 \ln [1/(|q| \al)]$ at $ q l_s \gtrsim 1 $,
$u_q$ and
$K_q$ depend logarithmically  on $q$ and
the excitations have a 1D plasmon-type spectrum $\om(q) \propto q \sqrt{\ln(1/ q)}$.
At spatial scales exceeding the screening length $l_s$, $q l_s \lesssim 1 $,
the interactions become effectively short-range with $V(q)$ saturating to the value $V(q \lesssim 1/l_s) = 2 e_*^2 \ln (l_s/\al)$.
The velocity $u_q=u$ and interaction parameter $K_q=K$
become  $q$-independent, $u/v=1/K= \sqrt{r_s \ln( l_s/ \al_*)}$,
and the spectrum $\om(q)=u |q|$ linear.
In the absence of the magnetic field
the spectrum is gapless, but for unscreened Coulomb  interactions the log-dependence of the velocity $u_q$ signals of a strong tendency towards gap opening.

Let us now study the correlations.
The operators that describe coupling between the counter-propagating helical modes are
given by the ``spin-flip'' components $s_\pm(x)=s_x(x) \pm \itxt s_y(x)$ of the spin density (\ref{eq:sxsy}),
\beq
    s_+(x) = \psi^\dg_\ua(x) \psi_\da(x) =
    \frac{ \etxt^{-2 \itxt \vphi(x)} }{2\pi \al} .
\label{eq:s+}
\eeq
The tendency towards gap opening is thus directly related
to the spin polarization in the $xy$ plane of the sample.
Calculating the correlation function of $s_\pm(x)$ with respect to the action (\ref{eq:S}) at zero temperature  $T=0$, we obtain
\beq
    \lan s_+(x) s_-(0) \ran
     \propto
     \lt\{ \begin{array}{l} \exp\lt[-4 \sqrt{\ln\lt(|x|/\al_*\rt)/r_s}\rt], \mbox{ } |x| \lesssim l_s, \\
            \lt(l_s/|x| \rt)^{2K}, \mbox{ } |x| \gtrsim l_s.
            \end{array} \rt.
\label{eq:ss}
\eeq
For screened Coulomb interactions at $|x| \gtrsim l_s$
the correlations (\ref{eq:ss}) of the in-plane spin density $s_{x,y}(x)$ have a LL power-law decay.
For unscreened Coulomb interactions at $|x| \lesssim l_s$,
the decay is  slower than any power law.
The interactions in the helical liquid
thus result in the tendency towards easy-plane ferromagnetic ordering. However, due to strong quantum fluctuations in a 1D system
the long-range order is not formed, $\lan \s(x) \ran = 0$.
For  unscreened Coulomb interactions, the tendency towards ferromagnetism
is  as strong as that towards Wigner crystallization in a conventional one-dimensional electron system~\cite{Schulz,FGS}.
Note that numerical factors in the spectrum $\om(q)=u_q |q|$~[Eq.~(\ref{eq:uK})] and correlation function (\ref{eq:ss}) differ from those of Refs.~\cite{Schulz,FGS} because in our case  electrons are single-flavored.

In the massless LL phase, the edge conductance  $G_\text{edge}=e^2/h$ is essentially unaffected by the interactions
and the edge  remains a perfect conducting channel~\cite{MS,SafiSchulz}.
Therefore, in the absence of perturbations that break time-reversal symmetry,
the interactions do not reveal themselves
in the transport measurement of either the two-terminal or Hall-bar longitudinal conductance $G=2 G_\text{edge} =2e^2/h$,
where the factor 2 is due to two edges in the former case and due to the mode equilibration in the contacts in the latter case.

\begin{figure}
$\mbox{\includegraphics[width=.27\textwidth]{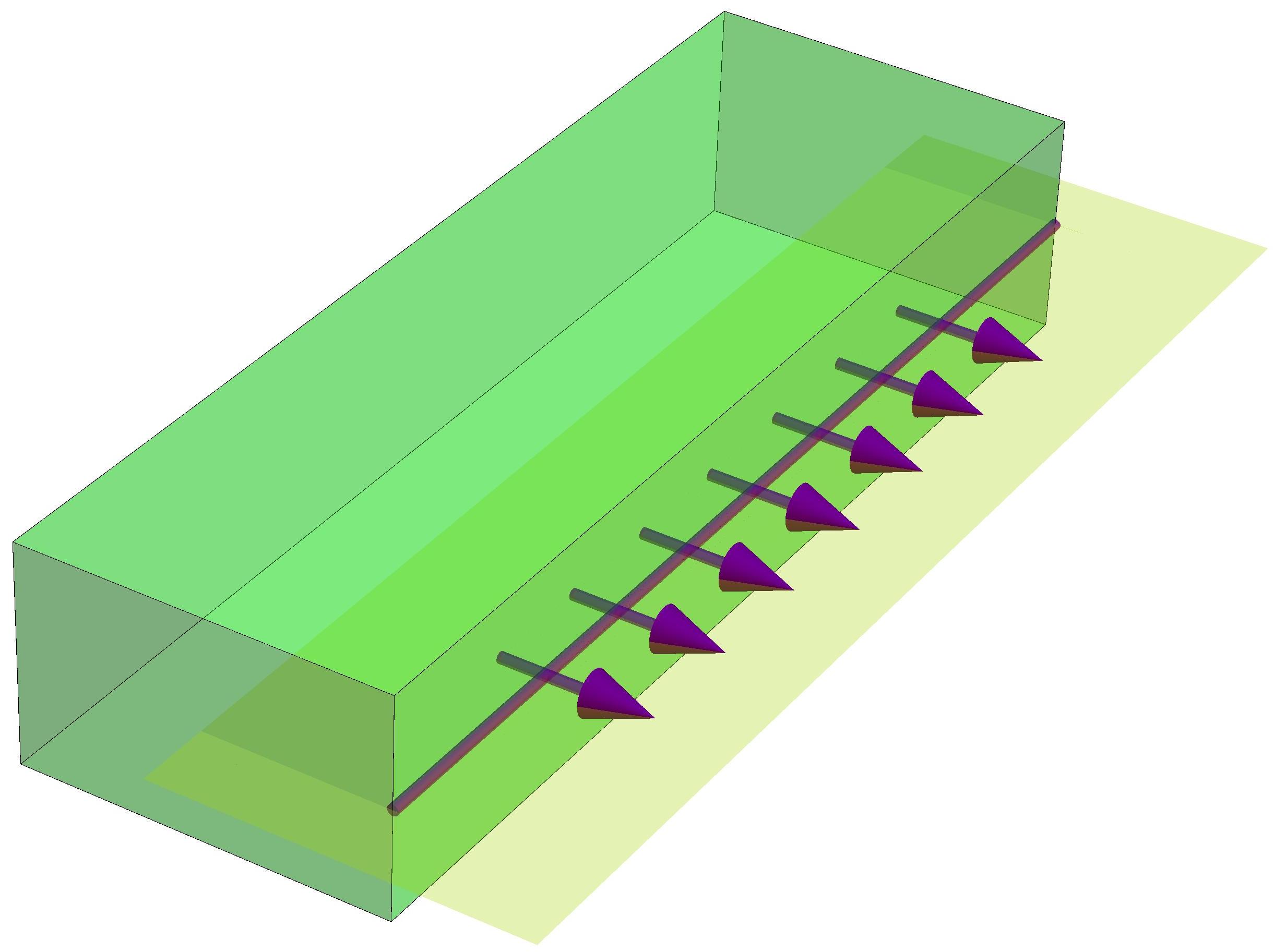}}\mbox{ }
\mbox{\includegraphics[width=.20\textwidth]{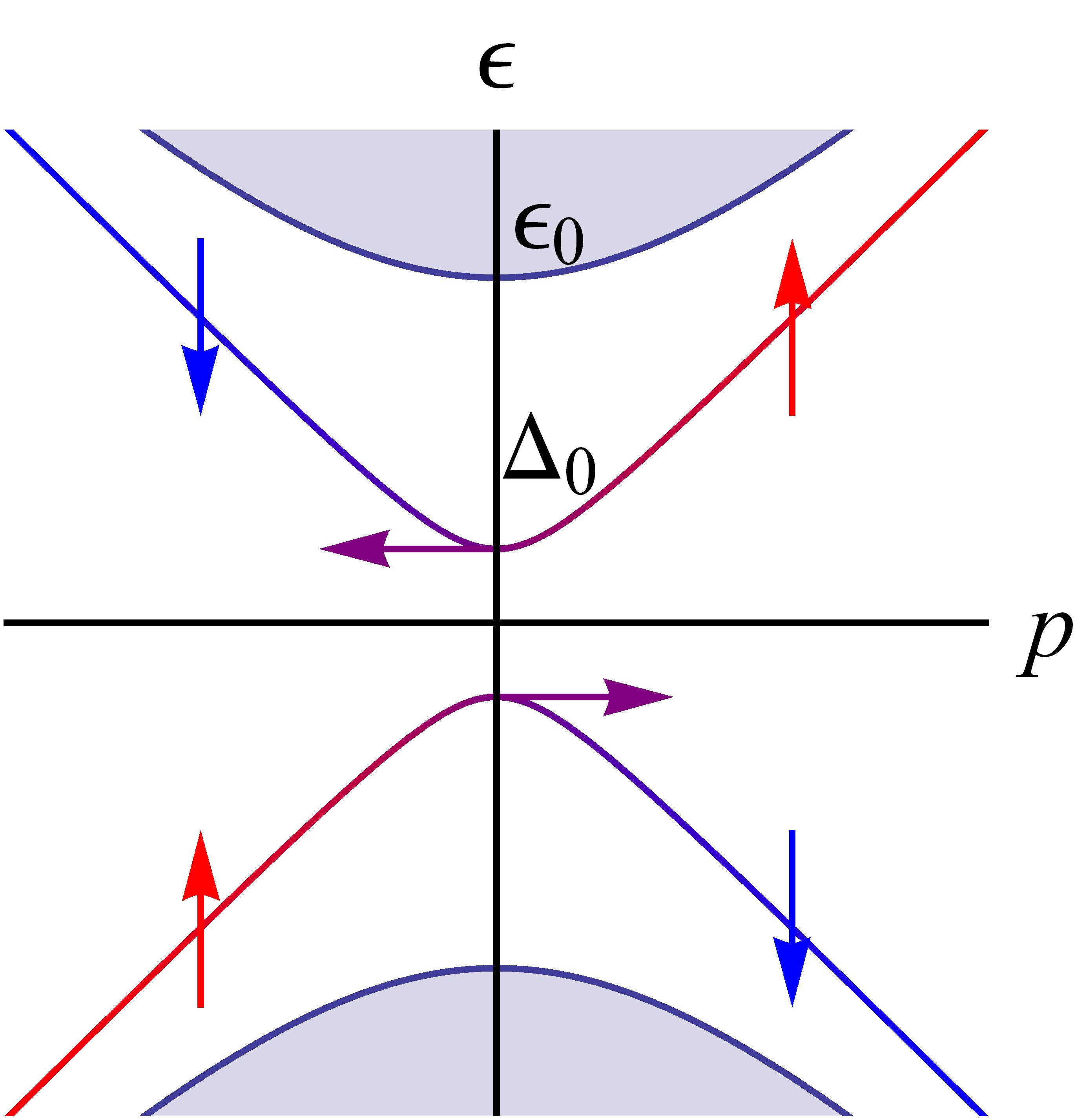}}$
\caption{
(Color online)
Magnetically ordered insulating state at the edge of a 2D topological insulator.
Magnetic field couples the spin up and down helical states, opens a gap $\De_0$ in the single-particle edge spectrum (right),
and polarizes the electron spins in the plane of the sample (left).
The many-body gap $\De$ [Eq.~(\ref{eq:De})]
is enhanced by the interactions
compared to the bare gap $\De_0$.
}
\label{fig:magnetic}
\end{figure}

\section{Gapped magnetically ordered phase at $B>0$}
The situation changes, if the magnetic field is applied, $\De_0 >0$ in $\Hh_\mtxt$ [Eq.~(\ref{eq:Hm})].
Even in the absence of interactions, the  magnetic field couples the helical counter-propagating states~\cite{HgTerev,sigrho} according to Eq.~(\ref{eq:Hm})
and opens a gap $\De_0$ in the single-particle spectrum $\e_p =\pm \sqrt{ (v p)^2 + \De_0^2 }$
of the Hamiltonian $\Hh_0+\Hh_\mtxt$, Fig.~\ref{fig:magnetic}.
In the ground state, the edge becomes spin polarized in the plane of the sample in the direction $\vphi_0$,
$\lan \s (x) \ran \propto (\cos \vphi_0,\sin \vphi_0,0)$.

Opening of the single-particle gap $\De_0$ has a direct consequence on transport.
For the noninteracting electrons, the edge conductance can be calculated using the Landauer formula
and for long enough edge of length $L\gg \hbar v/\De_0$ it is given by
\[
    G_\text{edge}(T) = \frac{2 e^2/h}{\exp(\De_0/T)+1}.
\]
The presence of the gap makes the edge insulating at temperatures $T \ll \De_0$, where
the conductance follows the Arrhenius activation law $G_\text{edge}(T) \approx 2(e^2/h) \exp(-\De_0/T)$.

Let us now take the interactions into account.
In terms of the collective excitations, the effect of the magnetic field is described by the cosine
term (\ref{eq:Hmb}) in the bosonized Hamiltonian.
The fact that the ground state is spin polarized
means that the phase field $\vphi(x)$ is locked in the minimum  of the cosine term, $\lan \vphi(x) \ran =-\vphi_0/2$.
The collective excitations are now massive and
for low energies described by the  fluctuations of $\vphi(x)$ around this minimum.
Since even without the magnetic field the interactions tend to order the edge ferromagnetically,
naturally, the gap $\De$ in the spectrum of  the collective excitations turns out to be enhanced compared to its bare single-particle value $\De_0$.
For screened Coulomb interactions we obtain
\beq
    \De\sim \e_0 \lt( \frac{\De_0}{\e_0} \rt)^{\frac{1}{2-K}} \propto B^{\frac{1}{2-K}},
\label{eq:De}
\eeq
up to a numerical factor $\sim 1$.
Here  $\e_0$ is the bulk insulator gap, which determines the high energy cutoff of the edge spectrum and is assumed  $\e_0 \gg \De_0$.
For HgTe quantum wells, it is estimated $\e_0 \sim 100 \text{K}$~\cite{HgTerev}.
The result (\ref{eq:De})
can be obtained by several means, e.g., using the self-consistent harmonic approximation~\cite{Giamarchi}.

The gap (\ref{eq:De}) has a power-law dependence on the bare gap $\De_0\sim \mu_B B$ and  hence
on the magnetic field $B$.
The exponent $1/(2-K)$ of this dependence is controlled by the LL interaction parameter $K$, which
varies between $K=1$ in the noninteracting case
and $K=0$ for infinitely strong
finite-range interactions; these cases give the lowest $\De_\text{min} = \De_0$ and highest  $\De_\text{max} \sim \sqrt{\De_0 \e_0} \propto \sqrt{B}$
possible values of the many-body gap $\De$, respectively.
Due to the long-range nature of the Coulomb forces, for unscreened interactions
 the gap appears to be close to $\De_\text{max}$ even for moderate interaction strength $r_s \sim 1$.
Performing the harmonic approximation~\cite{Giamarchi},
we obtain
\beq
    \De^2 \sim \De_0 \e_0  \exp[ -\sqrt{2 \ln (\e_0/\De_0) /r_s} ].
\label{eq:DeC}
\eeq
The gap (\ref{eq:DeC}) differs from the $K=0$ limit $\De_\text{max}$ of Eq.~(\ref{eq:De})
only by a function of $\De_0/\e_0$ that varies slower than any power law.
The result (\ref{eq:DeC}) applies if the correlation length
$l_\De = \hbar v/\De$  determined from Eq.~(\ref{eq:DeC}) does not exceed the screening length, $ l_\De \lesssim l_s$.
Otherwise, what concerns the gap, the interactions are effectively screened and
the gap is given by Eq.~(\ref{eq:De}).
For unscreened Coulomb interactions, the enhancement of the gap could thus be quite substantial:
for $\De_0 \sim 1 \text{K}$ and $\e_0 \sim 100 \text{K}$ one gets $\De_\text{max} \sim 10 \text{K}$.
The enhancement of the gap means, in particular, that interactions should favor observation of the
effects predicted in Ref.~\cite{sigrho}.

\section{Summary and experimental manifestation}
Summarizing, we studied the correlated magnetically ordered insulating state at the edge a of 2D topological insulator.
This spin-polarized state is induced by the application of the magnetic field and naturally facilitated  by electron interactions,
which drive the easy-plane ferromagnetic correlations in a helical liquid.
The key manifestation of the correlations is that the gap $\De \propto B^{1/(2-K)}$ [Eq.~(\ref{eq:De})]
in the spectrum of the collective spin-charge excitations exhibits a scaling dependence on the magnetic field $B$,
controlled by the Luttinger liquid parameter $K$, reflecting the quantum criticality of the helical liquid.

The main experimental implication of our findings is that electron interactions
should readily reveal themselves
in the insulating transport behavior of the magnetically ordered phase in a standard Hall-bar setup:
the gap $\De$ determines the activation dependence $G(T) \propto (e^2/h) \exp(-\De/T)$
of either two-terminal or longitudinal conductance at temperatures $T \ll \De$.
This should allow one to extract the Luttinger liquid parameter $K$ and infer about the strength of the interactions in the helical liquid
via the scaling dependence $\De \propto B^{1/(2-K)}$ of the gap.
Our findings thus suggest a Hall-bar device in an applied magnetic field as the minimal setup to
access the interaction-driven quantum criticality of the helical liquid at the edge of a 2D topological insulator.

\section{Acknowledgements}

Author is thankful to Konstantin  Matveev for valuable discussions.
This work was supported by the US DOE
under Contracts No. DE-AC02-06CH11357 and DE-FG02-99ER45790.

\end{document}